\documentclass{article}
\usepackage{spconf,amsmath,graphicx,hyperref}
\usepackage{algorithm}
\usepackage{algpseudocode}
\usepackage{xcolor}
\usepackage{amssymb}
\usepackage{svg}
\usepackage{multirow}
\usepackage{gensymb}
\usepackage{xurl}
\usepackage{hyperref}

\title{MarineSet: Large-Scale Underwater Acoustic Data Curation and evaluation}
\name{
Hilde I. Hummel$^{1}$ \qquad
Sandjai Bhulai$^{2}$ \qquad
Rob D. van der Mei$^{1,2}$ \qquad
Burooj Ghani$^{3}$
}

\address{
$^{1}$ Centrum Wiskunde \& Informatica, Amsterdam, The Netherlands\\
$^{2}$ Vrije Universiteit Amsterdam, Amsterdam, The Netherlands\\
$^{3}$ Naturalis Biodiversity Center, Leiden, The Netherlands
}
\begin{document}
%
\maketitle
\begin{abstract}

The scarcity of labeled underwater acoustic data limits the development of data-driven models for marine acoustic monitoring, motivating the use of self-supervised learning to exploit the abundance of unlabeled recordings. However, large-scale data alone does not guarantee effective representation learning, because model performance is strongly influenced by the distribution of the training data. To facilitate such training data, this study introduces a scalable curation pipeline that curates 3,000 hours from 41 years of raw NOAA hydrophone recordings, producing MarineSet, a large-scale and diverse dataset designed for underwater acoustic learning. A baseline model, MarineNet, trained on MarineSet achieves competitive performance on marine mammal call and ship-type classification, highlighting the value of systematic data curation for developing self-supervised learning models in underwater acoustics.

\end{abstract}
\begin{keywords}
Data Curation, Underwater Acoustics, Self-Supervised Learning
\end{keywords}
\section{Introduction}
\label{sec:intro}

\begin{figure}[htb]
\begin{minipage}[b]{1.0\linewidth}
  \centering
  \centerline{\includegraphics[width=8.5cm]{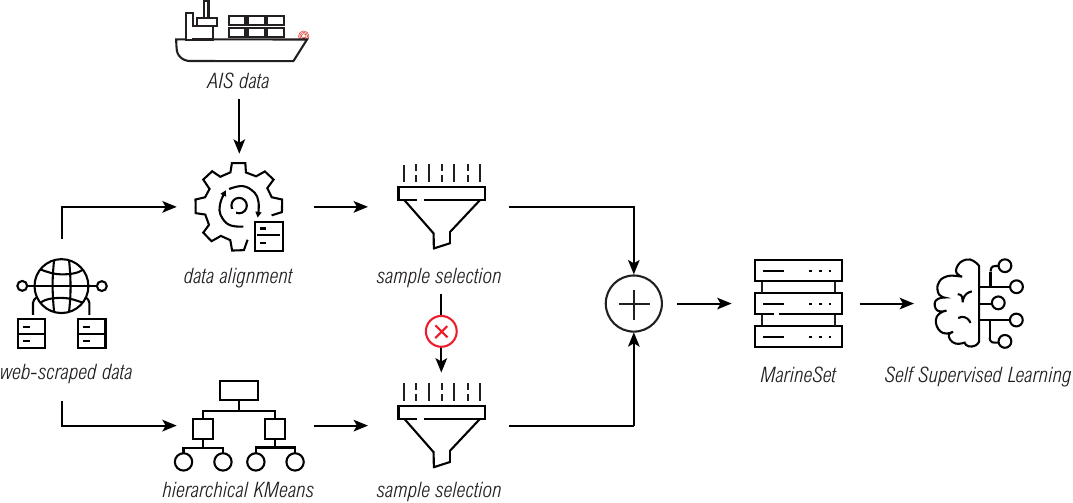}}
\end{minipage}
\caption{Overview of the proposed data curation pipeline.}
\label{fig:overview}
\end{figure}

The ocean has a diverse soundscape, however, the increasing level of anthropogenic noise threatens the sustainability of our ocean environment. This makes quantifying sound pollution and monitoring marine biodiversity crucial to protecting our ocean environment \cite{hummel2024survey}. For this purpose, various \textit{Passive Acoustic Monitoring} (PAM) systems are placed worldwide, recording various ocean soundscapes for a long period of time \cite{parsons2022sounding, NOAAurlPortal, ONCurlPortal}. The resulting large quantity of recordings creates the potential for automated analysis using \textit{Machine Learning} (ML) \cite{hummel2024survey}. While supervised learning has achieved the best performance in terrestrial bioacoustics \cite{van2025perch}, the scarcity of labeled underwater recordings makes this approach infeasible at the scale required for general-purpose models. \textit{Self-Supervised Learning} (SSL) has already shown promise in underwater acoustics \cite{li2025cross, xu2023self, xu2023selfmixup, hummel2026computation}, providing a viable approach for developing large foundation models from unlabeled recordings. However, these models were not trained on large, curated underwater acoustic datasets, despite the necessity of sufficient scale, diversity, and balance for SSL generalization \cite{vo2024automatic}. High-quality data curation is therefore essential \cite{xu2023demystifying, jose2025dinov2, hummel2025automated}. Motivated by this need, this study introduces \textit{MarineSet}, a large-scale curated dataset constructed from a substantial volume of web-scraped underwater acoustic recordings collected by hydrophones deployed across U.S. waters. The extensive data coverage of MarineSet enables the representation of a broad range of underwater soundscapes, providing a foundation for developing SSL models for automatic underwater acoustic analysis. To construct MarineSet, a previously introduced curation strategy \cite{hummel2025automated} is applied to a substantially larger volume of recordings. This strategy combines \textit{Automatic Identification System} (AIS) based vessel curation with unsupervised clustering-based sampling to obtain a balanced representation of the available recordings. An overview of the resulting pipeline is shown in Figure \ref{fig:overview}\footnote{Code and additional resources are available at \url{https://github.com/hildeingvildhummel/MarineSet}.}. To establish a baseline, \textit{MarineNet} is trained on the complete MarineSet dataset. MarineNet achieves competitive performance compared with existing underwater acoustic models, demonstrating the potential of MarineSet for SSL.

\section{Experimental setup}
\label{sec:methods}

\subsection{Web-scraped data}
\begin{figure}[htb]
\begin{minipage}[b]{1.0\linewidth}
  \centering
  \centerline{\includegraphics[width=8.5cm]{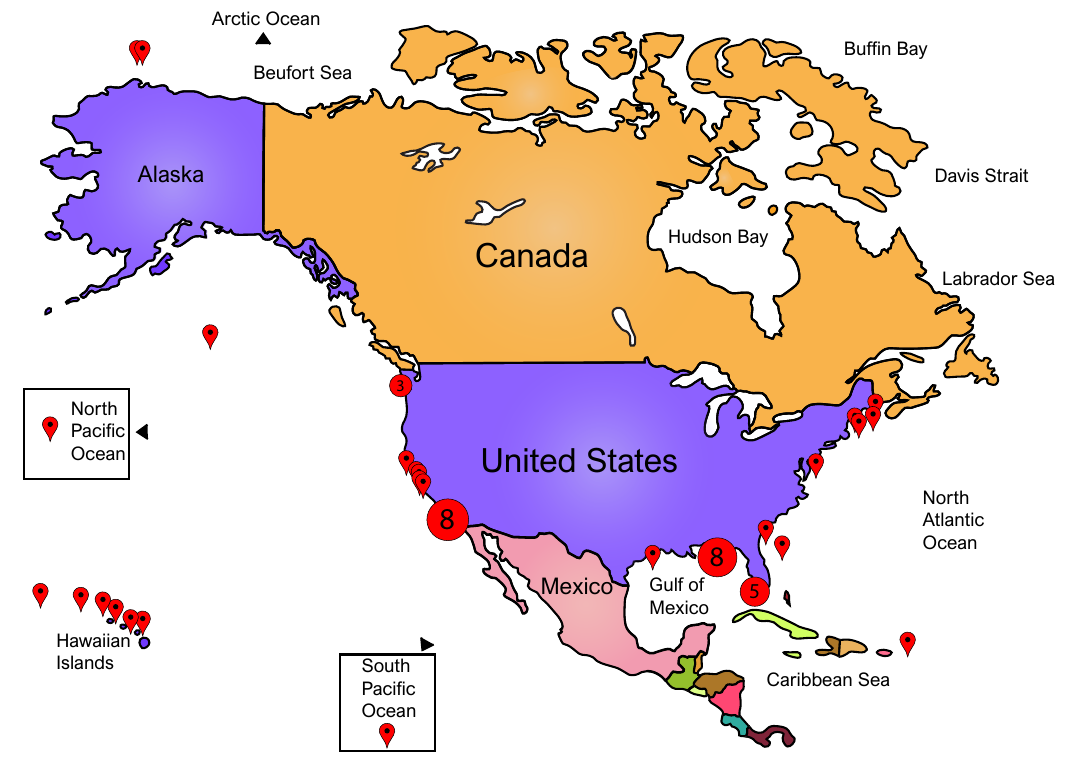}}
\end{minipage}
\caption{Hydrophone locations of the web-scraped data.}
\label{fig:expanded_hydrophones}
\end{figure}

\textit{National Oceanic and Atmospheric Administration} (NOAA) provides a large-scale repository of freely accessible underwater acoustic recordings from hydrophones distributed across U.S. waters \cite{NOAAurlPortal}. The geographic diversity of these recordings, spanning environments from Hawaii to the Arctic, provides a suitable basis for constructing a diverse curated dataset. From the complete archive, all hydrophones with recordings from 2019 onward were selected, resulting in 48 hydrophones (Figure~\ref{fig:expanded_hydrophones}) and a total duration of 41 years, 2 months, 11 days, 12 hours, 19 minutes, and 38 seconds. 

\subsection{AIS based data curation}
NOAA provides publicly available AIS data (\href{https://marinecadastre.gov/accessais/}{\textit{NOAA AIS source}}) covering U.S. territorial waters, enabling alignment with underwater acoustic recordings. Since \textit{ship-radiated noise} (SR-N) can exhibit low contrast with background noise, a clustering-based approach alone may fail to preserve vessel recordings. AIS information is therefore incorporated to identify vessel-present recordings and ensure vessel coverage during curation. As AIS data was unavailable for some locations, this process was limited to 28 hydrophones. For each hydrophone, a 4$\times$4 km recording range centered on the hydrophone was defined (similar to \cite{irfan2021deepship}), and all vessels transmitting an AIS message within this area were considered recorded. This alignment provides weak labels for curation, indicating that a vessel was within the defined range, without confirming that its acoustic signal was captured. This distinction is acceptable for curation, where the objective is to retain vessel diversity rather than assign precise acoustic labels. To reduce vessel imbalance, a duration threshold $t$ is applied per vessel. All recordings are retained for vessels with total duration below $t$, while recordings from vessels exceeding $t$ are sampled with probability inversely proportional to their total duration. The complete procedure is given in Algorithm~\ref{alg:AIS}. The resulting AIS-curated dataset is additionally annotated with vessel Direction of Arrival (DOA), estimated every 5~s by interpolating vessel positions between AIS transmissions.



\begin{algorithm}
\caption{AIS curation algorithm}\label{alg:AIS}
\begin{algorithmic}
\State \# $X$: raw PAM recordings
\State \# $A$: AIS data
\State \# $t$: threshold value
\State $X\_AIScurated = []$
\State $X\_ships$ = align\_AIS\_with\_audio($X, A$)
\State vessel\_count = extract\_occurrence\_unique\_vessels($X\_ships$)
\State vessel\_count[vessel\_count $< t$] = $t$
\State vessel\_prob = $t$ / vessel\_count
\For{vessel, audio in $X\_ships$}
\State vessel\_id = get\_index(vessel, vessel\_count)
\If{random.random() $<$ vessel\_prob[vessel\_id]}
    \State $X\_AIScurated$.append(audio)
    \State break
\EndIf
\EndFor
\end{algorithmic}
\end{algorithm}

\subsection{Clustering-based data curation}
The clustering-based curation of recordings without AIS information follows the approach proposed in \cite{vo2024automatic}. Since the web-scraped NOAA dataset already provides large-scale and diverse recordings through numerous hydrophones across different ocean environments, the clustering step focuses on balancing the data distribution. Samples are selected from an approximately uniform distribution $U$ over the support of the original distribution $P$. Raw recordings are embedded using the Conformer model from \cite{hummel2026computation} and clustered with hierarchical \textit{k}-means (HK-means), where each layer clusters the centroids of the previous layer. To prevent the exponential reduction in samples across layers, $r_t$ points closest to each centroid are resampled as input to the next layer. Repeated quantization progressively flattens the centroid distribution, approximating a uniform distribution \cite{vo2024automatic} (see Algorithm \ref{alg:hkmeans}). For NOAA curation, HK-means was configured with four layers and \textit{k} values in [6000, 400, 40, 10], with $r_t$ = [2200, 8, 5, 2] and $m=10$ resampling iterations. The model was trained on the complete web-scraped NOAA dataset using a streaming implementation to handle its scale \cite{sculley2010web}. Samples were selected through top-down hierarchical sampling, excluding samples that were already sampled during the AIS data curation. To approximate uniform sampling across clusters of different sizes, the number of selected samples per cluster is determined by finding the integer $n$ that minimizes $|N-\sum_j^k\min(n,s_j)|$, where $N$ is the target size of MarineSet and $s_j$ is the size of cluster $j$. This results in all samples being selected from smaller clusters and at most $n$ samples from larger clusters. The optimal $n$ is found using binary search, and samples closest to the centroids are selected. The target size $N$ was set to 3,000 hours minus the duration of the AIS-curated dataset. The AIS-curated dataset is combined with clustering-based curated dataset to define MarineSet.


\begin{algorithm}
\caption{Hierarchical \textit{k}-means with resampling originated from \cite{vo2024automatic}}\label{alg:hkmeans}
\begin{algorithmic}
\algnewcommand\Input{\item \textbf{Input: }} 
\algnewcommand\Result{\item \textbf{Result: }}
\Input Data $X \in \mathbb{R}^{nd\times d}$, number of levels $T$, number of clusters per level $(k_t)_{1 \leq t \leq T}$, number of resamplings $m$, number of points resampled per cluster $(r_t)_{1 \leq t \leq T}$.
\Result A hierarchy of clusters on data: centroids $(C_t)_{1 \leq t \leq T}$ and clusters $ ((L_t^{(i)})_{1 \leq i \leq k_t})_{1 \leq t \leq T}$.
\For{$t = 1$ to $T$}
    \State \textbf{if} $t=1$ \textbf{then} $\boldsymbol{I}\gets X$ \textbf{else} $\boldsymbol{I}\gets C_{t-1}$ 
    \State $C_t \gets kmeans(\boldsymbol{I}, k_t)$ 
    \State $L_T \gets assign(\boldsymbol{I}, C_t)$ 
    \State \# resampling-clustering
    \For{$s= 1 $ to $m$}
        \State $R \gets \cup_{i=1}^{k_t} resample (L_t^{(i)},r_t)$ 
        \State $C_t \gets kmeans(R,k_t)$ 
        \State $L_t \gets assign(\boldsymbol{I},C_t)$ 
    \EndFor
\EndFor
\end{algorithmic}
\end{algorithm}

\subsection{MarineNet: Baseline}
As a baseline, \textit{MarineNet} is trained on the complete MarineSet (AIS- and clustering-based curation) using a randomly initialized Wav2Vec~2.0 architecture \cite{baevski2020wav2vec}. The model is adapted to underwater acoustics through self-supervised training with the Wav2Vec~2.0 objective. It is evaluated on ShipsEar \cite{santos2016shipsear}, DeepShip \cite{irfan2021deepship}, Watkins \cite{sayigh2016watkins}, and DCLDE 2026 \cite{palmer2025public} using a frozen encoder and a linear sigmoid classifier trained with binary cross-entropy. Since no SSL model has been trained on a curated underwater acoustic dataset, MarineNet is compared with supervised models Google Whale \cite{googlewhale} and SurfPerch \cite{williams2024leveraging}. Performance is measured using mean Average Precision (mAP) and weighted ROC-AUC. 


\section{Results}
\label{sec:results}

\subsection{AIS based data Curation}
\begin{figure}[htb]

\begin{minipage}[b]{1.0\linewidth}
  \centering
  \centerline{\includegraphics[width=8.5cm]{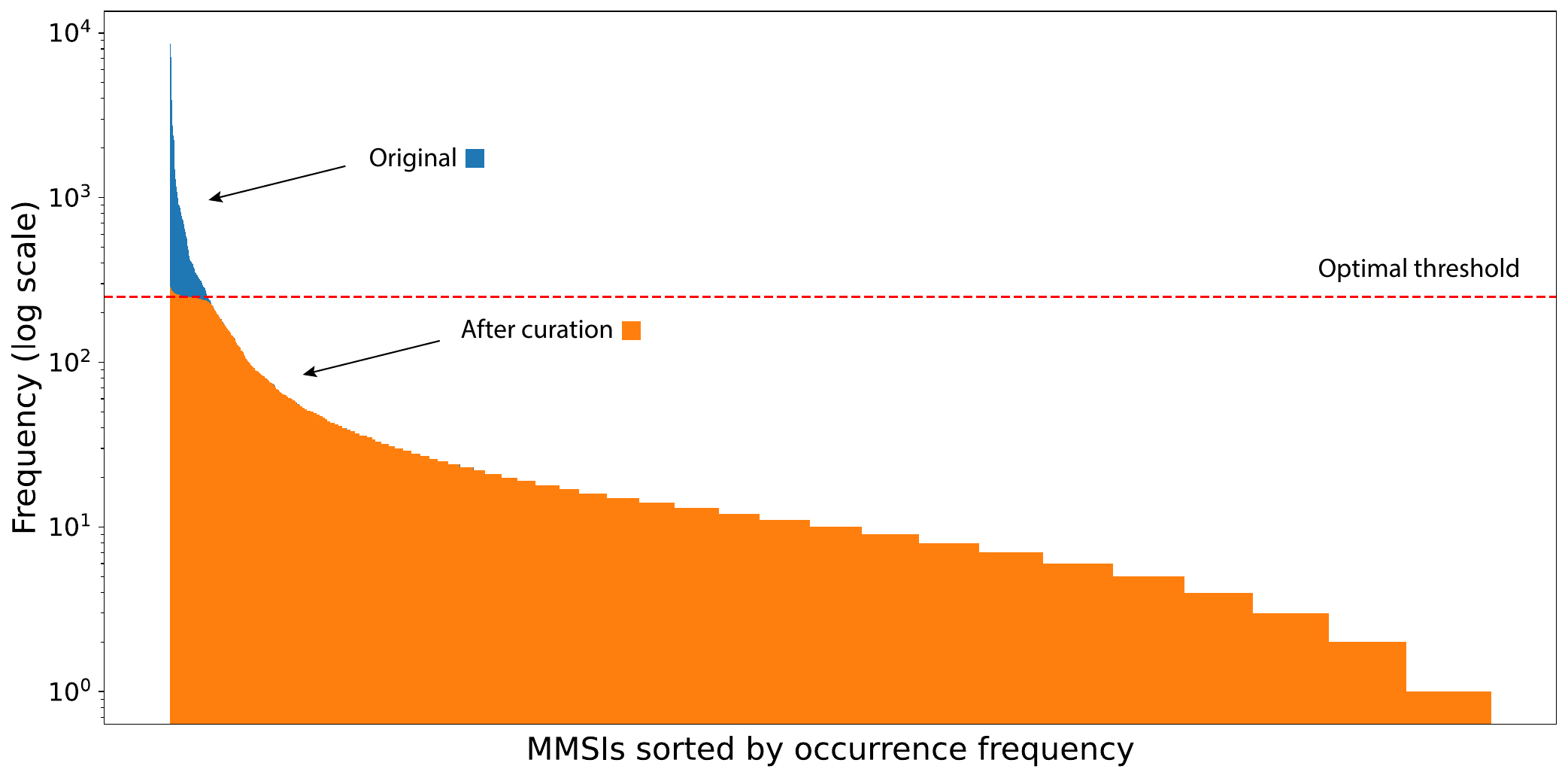}}
\end{minipage}
\caption{AIS aligned data distribution, with the original distribution in blue and the curated distribution in orange. $t_{opt}$ is visualized as a red line.}
\label{fig:Vessels}
\end{figure}

The distribution of unique MMSIs (Maritime Mobile Service Identity), corresponding to vessel IDs, before curation is shown in Figure~\ref{fig:Vessels}, revealing a strong imbalance in which a small number of vessels dominate the recordings. The optimal threshold-value $t$ was selected at the knee of this distribution, corresponding to $t=250$. After AIS-based curation, the dataset contains 968.2 hours of recordings from 6,540 unique MMSIs across 28 individual hydrophones. As shown in Figure~\ref{fig:Vessels}, the curation process substantially reduces the dominance of frequently recorded vessels, resulting in a more balanced vessel distribution while maintaining large-scale coverage and diversity.


\subsection{Clustering-based data curation}
\begin{figure}[htb]

\begin{minipage}[b]{1.0\linewidth}
  \centering
  \centerline{\includegraphics[width=8.5cm]{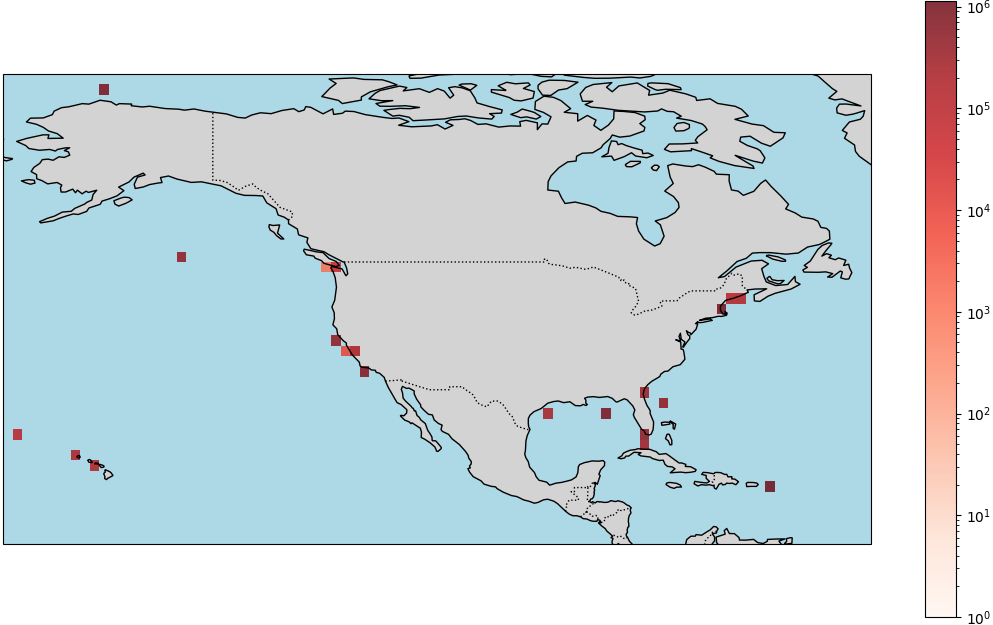}}
\end{minipage}
\caption{Geographical distribution of recordings selected through clustering-based curation. Colors indicate the total duration of selected recordings in seconds.}
\label{fig:MarineSet_map}
\end{figure}

\begin{figure}[htb]

\begin{minipage}[b]{1.0\linewidth}
  \centering
  \centerline{\includegraphics[width=8.5cm]{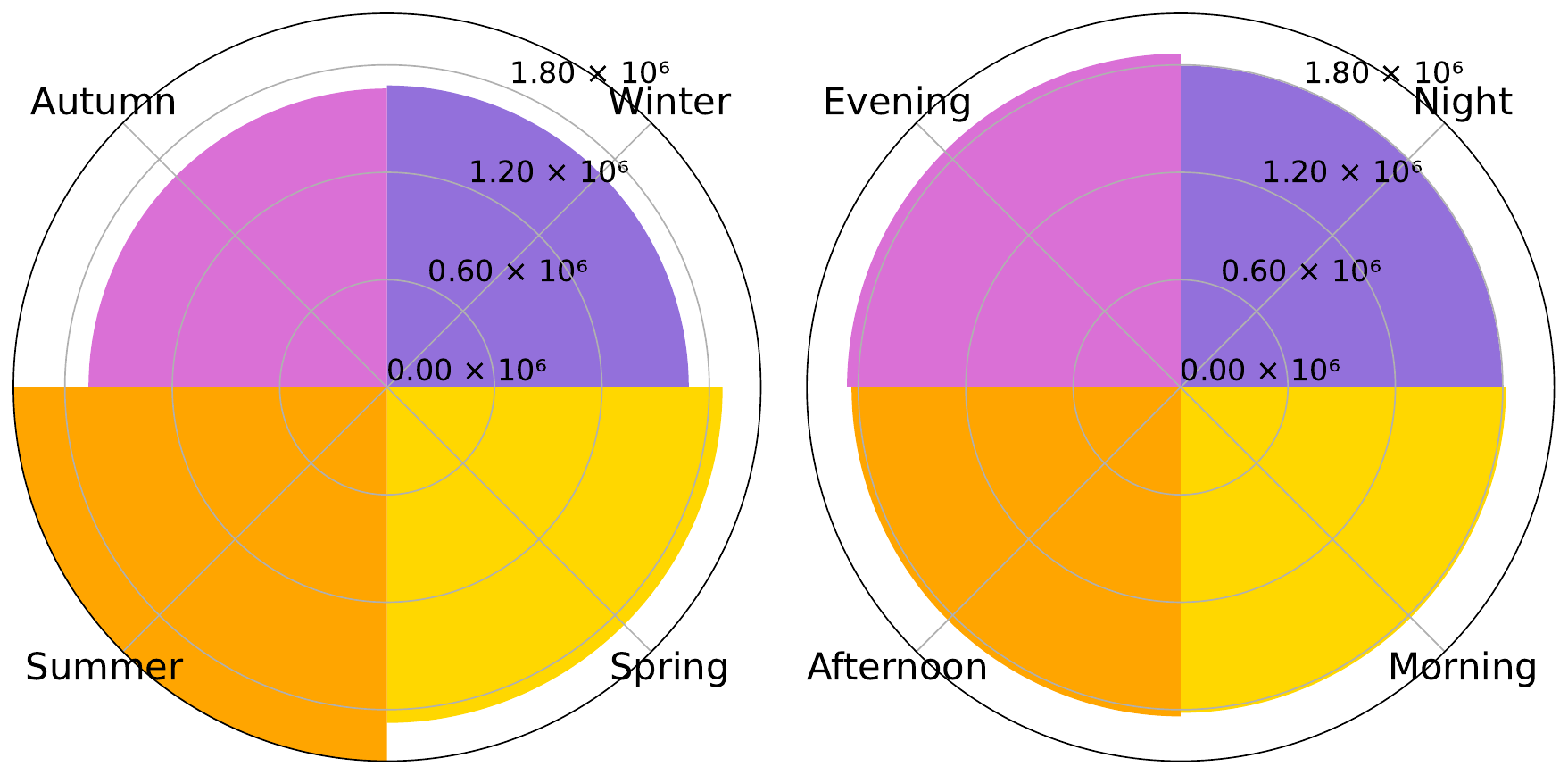}}
\end{minipage}
\caption{Distribution of recordings selected through clustering-based curation, with seasons shown on the left and time of day on the right. The size of each quarter represents the total duration of selected recordings in seconds.}
\label{fig:seasonality}
\end{figure}

The resulting curated dataset contains 2,031.8 hours of recordings from 43 hydrophones, geographically distributed across U.S. waters (Figure~\ref{fig:MarineSet_map}). In contrast to the original geographical distribution (Figure~\ref{fig:expanded_hydrophones}), previously overrepresented areas are no longer dominant, resulting in a more balanced geographic distribution. The hydrophones span diverse environments, from the Arctic to the Gulf of Mexico, preserving broad marine acoustic diversity despite the rebalancing process. Seasonal and diurnal variation is also represented (Figure~\ref{fig:seasonality}), with recordings covering all four seasons and an approximately uniform distribution across the day, with summer being slightly overrepresented.

\subsection{Underwater Acoustic Classification}
To evaluate MarineNet as a baseline model, its performance is assessed on ship-type classification and on marine mammal call classification (see Table \ref{tab:classification}). MarineNet demonstrates competitive performance relative to supervised models, outperforming Google Whale on both DeepShip and Watkins. However, both Google Whale and MarineNet show limited performance on the multilabel classification task represented by DCLDE2026. SurfPerch consistently achieves superior performance, which is expected given that it was trained on a large, labeled underwater acoustic dataset specifically designed for marine life classification.


\begin{table}[]
    \centering
    \begin{tabular}{c c c c c c c}
        \multirow{ 2}{*}{Dataset} & \multicolumn{2}{c}{Google Whale} & \multicolumn{2}{c}{SurfPerch} & \multicolumn{2}{c}{MarineNet}  \\
        & mAP & AUC & mAP & AUC & mAP & AUC \\
        \hline
         ShipsEar & 0.56 & 0.75 & \textbf{0.62} & \textbf{0.82} & 0.54 & 0.74 \\
         Deepship & 0.46 & 0.74 & \textbf{0.60} & \textbf{0.82} & 0.55 & 0.78 \\
         \hline
         Watkin's & 0.55 & 0.96 & \textbf{0.89} & \textbf{0.99} & 0.65 & 0.96 \\
         DCLDE 2026 \\ Species & 0.53 & 0.86 & \textbf{0.76} & \textbf{0.95} & 0.42 & 0.81 \\
    \end{tabular}
    \caption{Classification performance underwater on acoustic tasks.}
    \label{tab:classification}
\end{table}

\section{Discussion}
The proposed data curation pipeline demonstrates how a large, diverse, and balanced dataset can be extracted from massive volumes of web-scraped underwater acoustic data. Through unsupervised clustering, the method captures geographic and seasonal variability without requiring manual annotations. The resulting MarineSet captures a broad range of underwater acoustic soundscapes across U.S. waters. This diversity extends beyond existing datasets, which often focus on specific marine mammal species or contain limited amounts of ship-radiated noise data, making MarineSet suitable for a broad range of underwater acoustic applications. MarineSet also provides a foundation for self-supervised learning (SSL), as demonstrated by the competitive performance of the baseline MarineNet model on both ship-type classification and marine mammal call classification tasks. While our baseline MarineNet model demonstrates the viability of SSL on MarineSet, the dataset itself, not the particular model, is the primary contribution. MarineNet serves as a proof-of-concept that curated, diverse, unsupervised data can support competitive underwater acoustic modeling. Although supervised approaches are generally known to outperform SSL-based models in bioacoustics \cite{van2025perch}, MarineNet achieves performance approaching that of supervised underwater acoustic models, despite serving just as a baseline model. This highlights the potential of MarineSet to support the development of SSL approaches for a diverse range of underwater acoustic tasks, including marine mammal and ship monitoring, underwater sound pollution assessment, and biodiversity monitoring. Future research will focus on the development of an optimized SSL model on MarineSet in continuation on the baseline MarineNet presented.





\section{Conclusion}
Underwater acoustics suffers from a scarcity of labeled datasets, limiting supervised learning for marine monitoring. Self-supervised learning (SSL) offers an alternative, but its application to underwater sound classification has been limited by the lack of suitable large-scale datasets. This work addresses this gap with an unsupervised curation pipeline that produces MarineSet, a large, diverse, and balanced underwater acoustic dataset designed for SSL. MarineNet, trained on MarineSet, achieves competitive performance with supervised approaches on ship-type and marine mammal call classification. These results demonstrate that curated unlabeled data can enable effective SSL for underwater acoustic monitoring without costly manual annotation. MarineSet provides a foundation for developing robust SSL models for applications including marine sound pollution assessment and biodiversity monitoring.

\bibliographystyle{IEEEbib}
\bibliography{strings,refs}

@article{santos2016shipsear,
  title={ShipsEar: An underwater vessel noise database},
  author={Santos-Dom{\'\i}nguez, D. and Torres-Guijarro, S. and Cardenal-L{\'o}pez, A. and Pena-Gimenez, A.},
  journal={Applied Acoustics},
  volume={113},
  pages={64--69},
  year={2016},
  month = {December}
}

@article{irfan2021deepship,
  title={DeepShip: An underwater acoustic benchmark dataset and a separable convolution based autoencoder for classification},
  author={Irfan, M. and Jiangbin, Z. and Ali, S. and Iqbal, M. and Masood, Z. and Hamid, U.},
  journal={Expert Systems with Applications},
  volume={183},
  pages={115270},
  year={2021},
  month={November}
}

@inproceedings{xu2023demystifying,
  title={Demystifying clip data},
  author={Xu, H. and Xie, S. and Tan, X.E. and Huang, P. and Howes, R. and Sharma, V. and Li, S. and Ghosh, G. and Zettlemoyer, L. and Feichtenhofer, C.},
  booktitle={Proc. International Conference on Learning Representations},
  year={2024},
  pages = {47812-47831}
}

@article{vo2024automatic,
  title={Automatic data curation for self-supervised learning: A clustering-based approach},
  author={Vo, H.V. and Khalidov, V. and Darcet, T. and Moutakanni, T. and Smetanin, N. and Szafraniec, M. and Touvron, H. and others},
  journal={arXiv preprint arXiv:2405.15613},
  year={2024}
}

@inproceedings{jose2025dinov2,
  title={Dinov2 meets text: A unified framework for image-and pixel-level vision-language alignment},
  author={Jose, C. and Moutakanni, T. and Kang, D. and Baldassarre, F. and Darcet, T. and Xu, H. and Li, D. and others},
  booktitle={Proc. Computer Vision and Pattern Recognition Conference},
  pages={24905--24916},
  year={2025}
}

@inproceedings{hummel2025automated,
  author    = {Hummel, H.I. and Bhulai, S. and Ghani, B. and van der Mei, R.D.},
  title     = {Automated Data Curation for Self-Supervised Learning in Underwater Acoustic Analysis},
  booktitle = {Proc. 11th Convention of the European Acoustics Association: Forum Acusticum / EuroNoise 2025},
  year      = {2025},
  month     = {May},
  pages={2165-2170}
}

@inproceedings{sayigh2016watkins,
  title={The Watkins marine mammal sound database: an online, freely accessible resource},
  author={Sayigh, L. and Daher, M. and Allen, J. and Gordon, H. and Joyce, K. and Stuhlmann, C. and Tyack, P.},
  booktitle={Proc. of Meetings on Acoustics},
  volume={27},
  pages={040013},
  year={2017},
  month={February}
}

@misc{ONCurlPortal,
author = {Ocean Network Canada},
title = {Dataportal},
howpublished = {\url{https://data.oceannetworks.ca/home}},
note = {Accessed: 9 December 2025},
year = {2007}
}

@misc{NOAAurlPortal,
author = {National Oceanic and Atmospheric Administration},
title = {Passive-Acoustic-Data},
howpublished = {\url{https://www.ncei.noaa.gov/products/passive-acoustic-data}},
note = {Accessed: 9 December 2025},
year = {2017}
}

@article{hummel2024survey,
  title={A survey on machine learning in ship radiated noise},
  author={Hummel, H.I. and van der Mei, R.D. and Bhulai, S.},
  journal={Ocean Engineering},
  volume={298},
  pages={117252},
  year={2024},
  month={April}
}

@article{hummel2026computation,
  title={The computation of generalized embeddings for underwater acoustic target recognition using contrastive learning},
  author={Hummel, H.I. and Gansekoele, A. and Bhulai, S. and van der Mei, R.D.},
  journal={Applied Acoustics},
  volume={243},
  pages={111103},
  year={2026},
  month={February}
}

@article{xu2023self,
  title={Self-supervised learning--based underwater acoustical signal classification via mask modeling},
  author={Xu, K. and Xu, Q. and You, K. and Zhu, B. and Feng, M. and Feng, D. and Liu, B.},
  journal={The Journal of the Acoustical Society of America},
  volume={154},
  number={1},
  pages={5--15},
  year={2023},
  month={July}
}

@article{xu2023selfmixup,
  title={Self-supervised learning-for underwater acoustic signal classification with mixup},
  author={Xu, Q. and Jiang, J. and Xu, K. and Dou, Y. and Gao, C. and Zhu, B. and You, K. and Zhu, Q.},
  journal={IEEE Journal of Selected Topics in Applied Earth Observations and Remote Sensing},
  volume={17},
  pages={3530--3542},
  year={2023},
  month={October}
}

@article{parsons2022sounding,
  title={Sounding the call for a global library of underwater biological sounds},
  author={Parsons, M.J.G. and Lin, T. and Mooney, T.A. and Erbe, C. and Juanes, F. and Lammers, M. and Li, S. and others},
  journal={Frontiers in Ecology and Evolution},
  volume={10},
  pages={810156},
  year={2022},
  month={February}
}

@inproceedings{sculley2010web,
  title={Web-scale k-means clustering},
  author={Sculley, D.},
  booktitle={Proc. of the 19th international conference on World wide web},
  pages={1177--1178},
  year={2010}
}

@inproceedings{baevski2020wav2vec,
  title={wav2vec 2.0: A framework for self-supervised learning of speech representations},
  author={Baevski, A. and Zhou, Y. and Mohamed, A. and Auli, M.},
  booktitle={Proc. 34th Conference on Neural Information Processing Systems},
  volume={33},
  pages={12449--12460},
  year={2020}
}

@article{williams2024leveraging,
  title={Leveraging tropical reef, bird and unrelated sounds for superior transfer learning in marine bioacoustics},
  author={Williams, B. and van Merri{\"e}nboer, B. and Dumoulin, V. and Hamer, J. and Triantafillou, E. and Fleishman, A.B. and McKown, M. and Munger, J.E. and others},
  journal={arXiv preprint arXiv:2404.16436},
  year={2024}
}

@misc{googlewhale,
  author = {Harrell, L.},
  title = {Whistles, songs, boings, and biotwangs: Recognizing whale vocalizations with AI},
  howpublished = {\url{https://research.google/blog/whistles-songs-boings-and-biotwangs-recognizing-whale-vocalizations-with-ai/}},
  year = {2024}, 
  note = {Accessed: 7-Oct-2025}
}

@article{palmer2025public,
  title={A public dataset of annotated orcinus orca acoustic signals for detection and ecotype classification},
  author={Palmer, K.J. and Cummings, E. and Dowd, M.G. and Frasier, K. and Frazao, F. and Harris, A. and Houweling, A. and others},
  journal={Scientific data},
  volume={12},
  number={1},
  pages={1137},
  year={2025},
  month={July}
}

@article{li2025cross,
  title={A cross-architecture masked contrastive learning framework for few-shot underwater acoustic target classification},
  author={Li, W. and Chen, Z. and Wang, J. and Liu, T. and Chen, H. and Xu, G.},
  journal={Knowledge-Based Systems},
  pages={114252},
  year={2025},
  month={October}
}

@article{van2025perch,
  title={Perch 2.0: The bittern lesson for bioacoustics},
  author={van Merri{\"e}nboer, Bart and Dumoulin, Vincent and Hamer, Jenny and Harrell, Lauren and Burns, Andrea and Denton, Tom},
  journal={arXiv preprint arXiv:2508.04665},
  year={2025}
}

\end{document}